\documentclass[aps,prd,twocolumn,floatfix,preprintnumbers,showpacs]{revtex4}
\usepackage{amssymb,amsmath,bm,graphicx,longtable}
\usepackage{mathrsfs,enumerate}

\allowdisplaybreaks

\begin{document}
\preprint{CHIBA-EP-193/KEK Preprint 2012-9}

\title{Magnetic monopole loops generated from two-instanton solutions: \\
Jackiw-Nohl-Rebbi  versus  't Hooft instanton}

\author{Nobuyuki Fukui$^{1}$}
\email[]{n.fukui@graduate.chiba-u.jp}

\author{Kei-Ichi Kondo$^{1}$}
\email[]{kondok@faculty.chiba-u.jp}

\author{Akihiro Shibata$^{2}$}
\email[]{akihiro.shibata@kek.jp}

\author{Toru Shinohara$^{1}$}
\email[]{sinohara@graduate.chiba-u.jp}

\affiliation{
$^1$Department of Physics, Graduate School of Science, Chiba University,
Chiba 263-8522, Japan
\\
$^2$Computing Research Center,
High Energy Accelerator Research Organization (KEK) 
and Graduate Univ. for Advanced Studies (Sokendai),
Tsukuba  305-0801, Japan
}

\date{\today}

\begin{abstract}
In our previous paper \cite{FKSS10}, we have shown that
the Jackiw-Nohl-Rebbi two-instanton generates a circular loop of magnetic monopole in the four-dimensional Euclidean SU(2) Yang-Mills theory. 
On the other hand, it is claimed in \cite{BOT97} that the 't Hooft two-instanton does not generate magnetic monopole loop.
It seems that two results are inconsistent with each other, since  the JNR two-instanton converges to the 't Hooft two-instanton in a certain limit.
In this paper, we clarify that two results are compatible with each other by demonstrating how the magnetic monopole loop generated from the JNR two-instanton  deforms in the process of taking the 't Hooft two-instanton limit.
\end{abstract}


\maketitle

\section{Introduction}\label{sec:intro}

It is believed that the dual superconductivity proposed in \cite{Nambu74,tHooft1975,Mandelsta76,Polyakov75} is a promising mechanism for quark confinement.
In this mechanism, condensation of magnetic monopoles causes confinement.
For this mechanism to work, therefore, it must be shown that magnetic monopoles to be condensed indeed exist in Yang-Mills theory. In the lattice simulation \cite{SKKMSI07b}, magnetic monopoles are shown to exist in the Yang-Mills theory and magnetic monopole currents form closed loops.

From this point of view, it is interesting to investigate whether or not   magnetic monopoles are generated from instantons.  
In fact, Brower,  Orginos and  Tan (BOT) \cite{BOT97} have investigated  numerically  whether or not magnetic monopoles are generated from the 't Hooft  instantons.  
Then they have shown that two magnetic monopole loops are generated from the 't Hooft two-instanton around two poles of the 't Hooft two-instanton, while one-instanton does not work for generating magnetic monopoles.  
However, they are localized in the neighborhood of two poles and hence the generation of magnetic monopoles is regarded as a lattice artifact, in other words, the 't Hooft two-instanton does not generate magnetic monopole loops in the continuum limit.

In our previous study \cite{FKSS10}, on the other hand,   we have shown that the Jackiw-Nohl-Rebbi (JNR) two-instanton \cite{JNR77} generates a circular loop of magnetic monopole.
Since the JNR instanton reduces to the 't Hooft instanton in a certain limit which we call the 't Hooft instanton limit, it seems that our result is inconsistent with the BOT result.

In this paper, we show that this inconsistency is resolved by examining more carefully the behavior of the magnetic monopole generated from the JNR two-instanton.
First, we specify numerically  the magnetic monopole generated from the JNR two-instanton by a similar procedure to that described in the previous paper \cite{FKSS10}.  
Then we investigate how  the magnetic monopole behaves in taking the 't Hooft instanton limit.
We demonstrate that a magnetic monopole loop is deformed  into two smaller loops and they eventually shrink to two poles of the 't Hooft instanton in the course of taking the 't Hooft limit.
As a result, we confirm that two results are compatible with each other.

\section{JNR instanton in 't Hooft instanton limit}

The JNR two-instanton configuration is given by
\begin{align}
 g{\bf A}_\mu^\text{JNR}(x)&=2T_A
                  \eta_{\mu\nu}^{A(-)}
                  \phi_\text{JNR}^{-1}
                  \sum_{r=0}^2\frac{\rho_r^2\left(x^\nu-b^\nu_r\right)}
                                   {|x-b_r|^4} , 
\\
 \phi_\text{JNR}&:=\sum_{r=0}^2\frac{\rho_r^2}{|x-b_r|^2}.
\end{align}
It is specified by three pole positions
$b_0=(b_0^1,b_0^2,b_0^3,b_0^4)$, $b_1=(b_1^1,b_1^2,b_1^3,b_1^4)$, $b_2=(b_2^1,b_2^2,b_2^3,b_2^4)$ 
and three size parameters $\rho_0,\rho_1,\rho_2$.
The limit of the JNR two-instanton as $|b_0|=\rho_0\rightarrow\infty$ is just the 't Hooft two-instanton having two poles
$(b_1^1,b_1^2,b_1^3,b_1^4)$, $(b_2^1,b_2^2,b_2^3,b_2^4)$:
\begin{align}
  g{\bf A}_\mu^\text{'t Hooft}(x)&=2T_A
                  \eta_{\mu\nu}^{A(-)}
                  \phi_\text{'t Hooft}^{-1}
                  \sum_{r=1}^2\frac{\rho_r^2\left(x^\nu-b^\nu_r\right)}
                                   {|x-b_r|^4} , \\
 \phi_\text{'t Hooft}&:=1+\sum_{r=1}^2\frac{\rho_r^2}{|x-b_r|^2}.
\end{align} 
Thus, the 't Hooft two-instanton is reproduced from the JNR two-instanton
as the location of the first
pole $b_0$ is sent to infinity keeping the relation $|b_0|=\rho_0$.

\vskip -1.0cm
\begin{figure*}
 \unitlength=0.001in
 \begin{picture}(7000,9000)(0,0)
  \put(0,6750){\includegraphics[trim=0 0 0 0, width=80mm]%
              {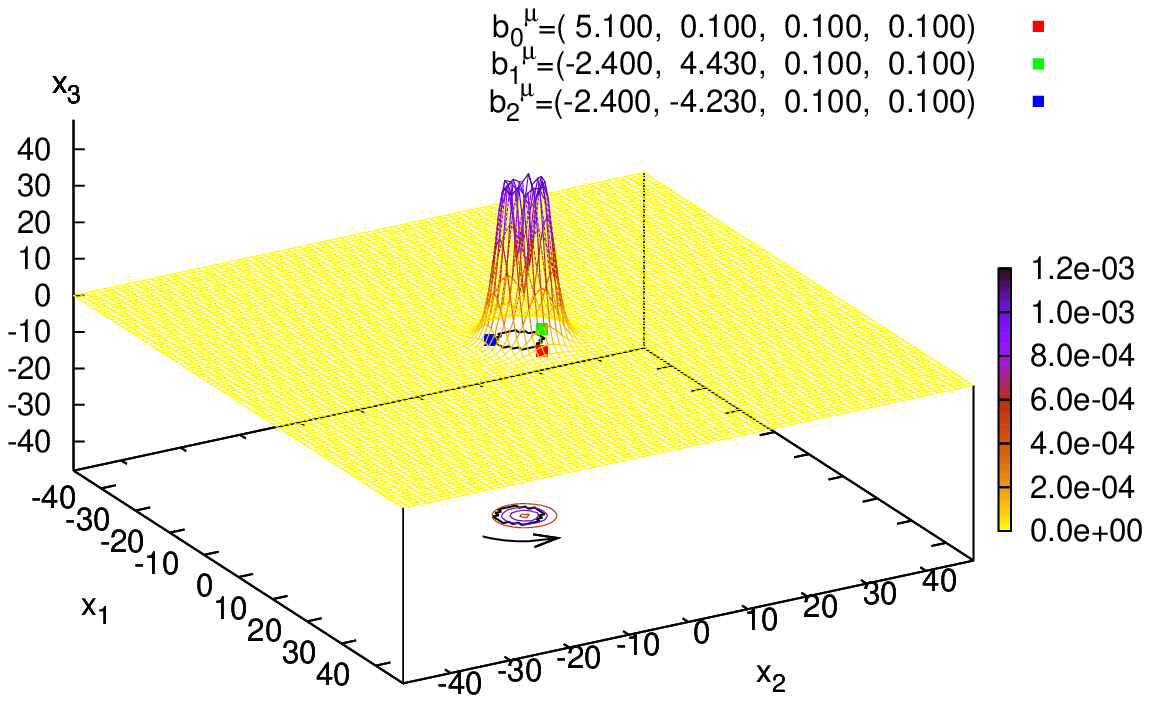}}%
  \put(1400,6750){(a)}
  \put(3500,6750){\includegraphics[trim=0 0 0 0, width=80mm]%
                 {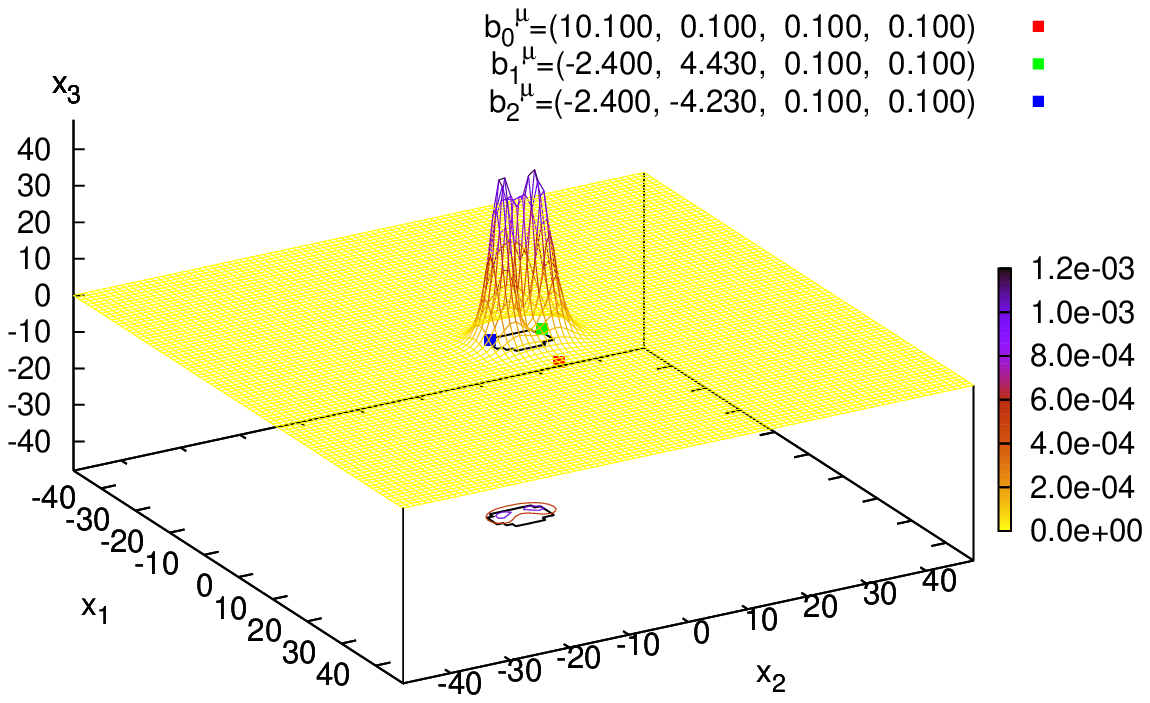}}%
  \put(4900,6750){(b)}
  \put(0,4500){\includegraphics[trim=0 0 0 0, width=80mm]%
           {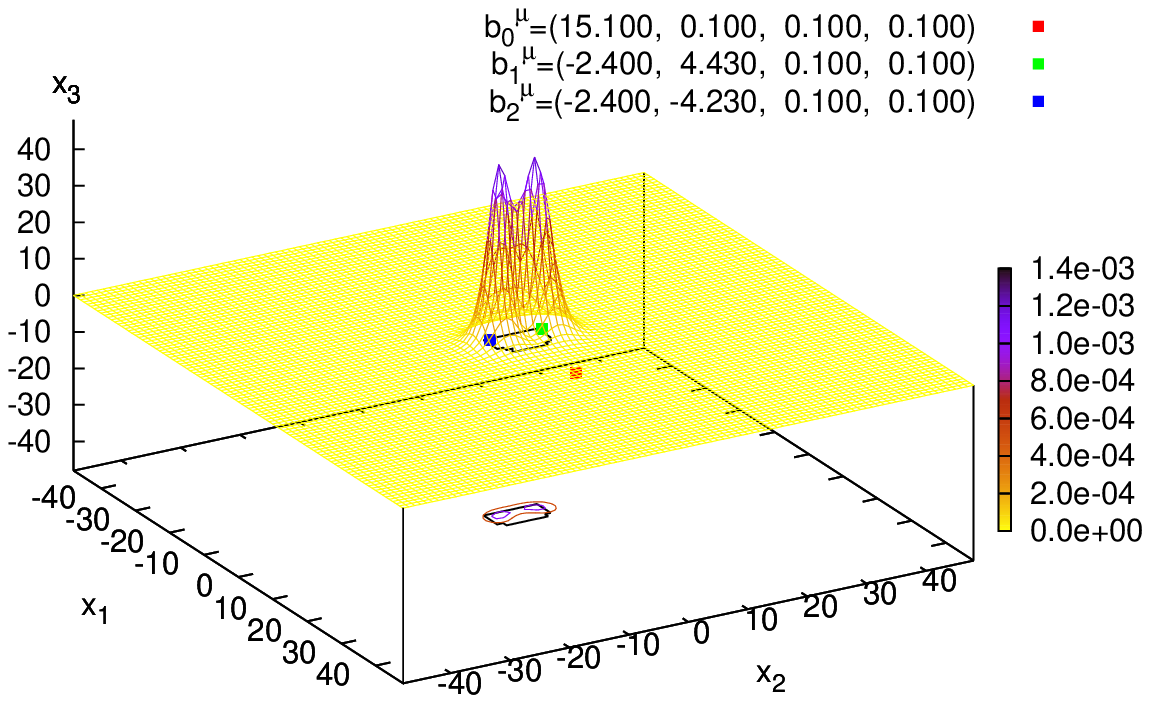}}%
  \put(1400,4500){(c)}
  \put(3500,4500){\includegraphics[trim=0 0 0 0, width=80mm]%
              {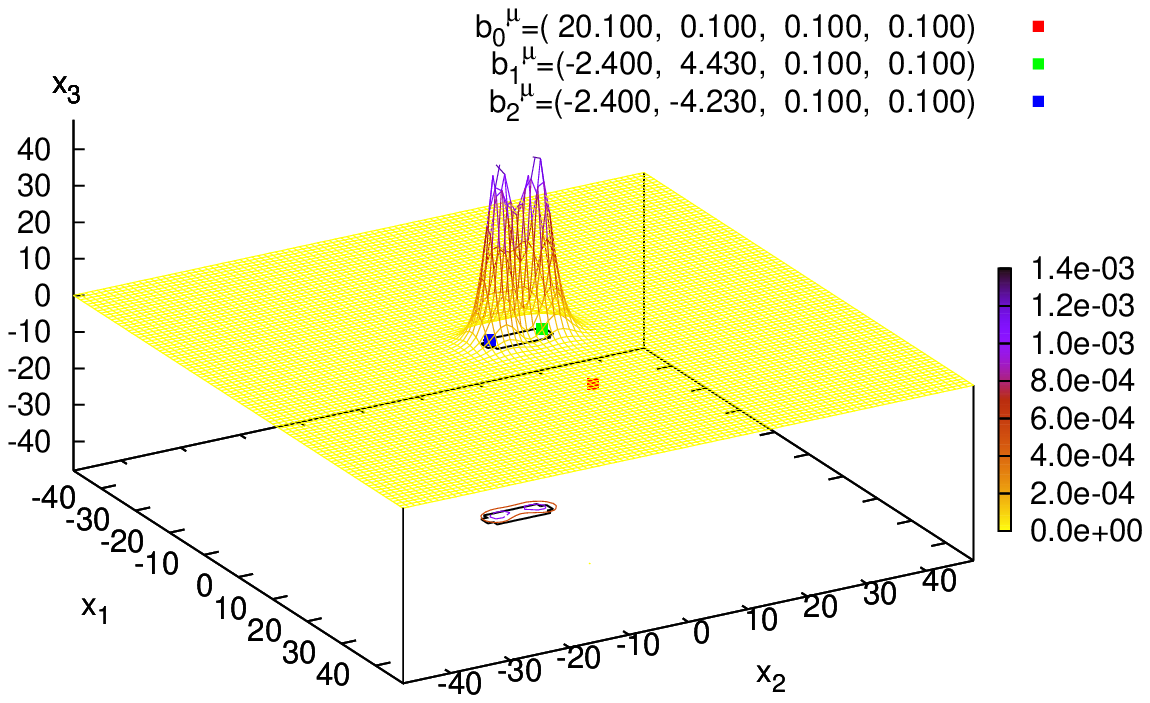}}%
  \put(4900,4500){(d)}
  \put(0,2250){\includegraphics[trim=0 0 0 0, width=80mm]%
              {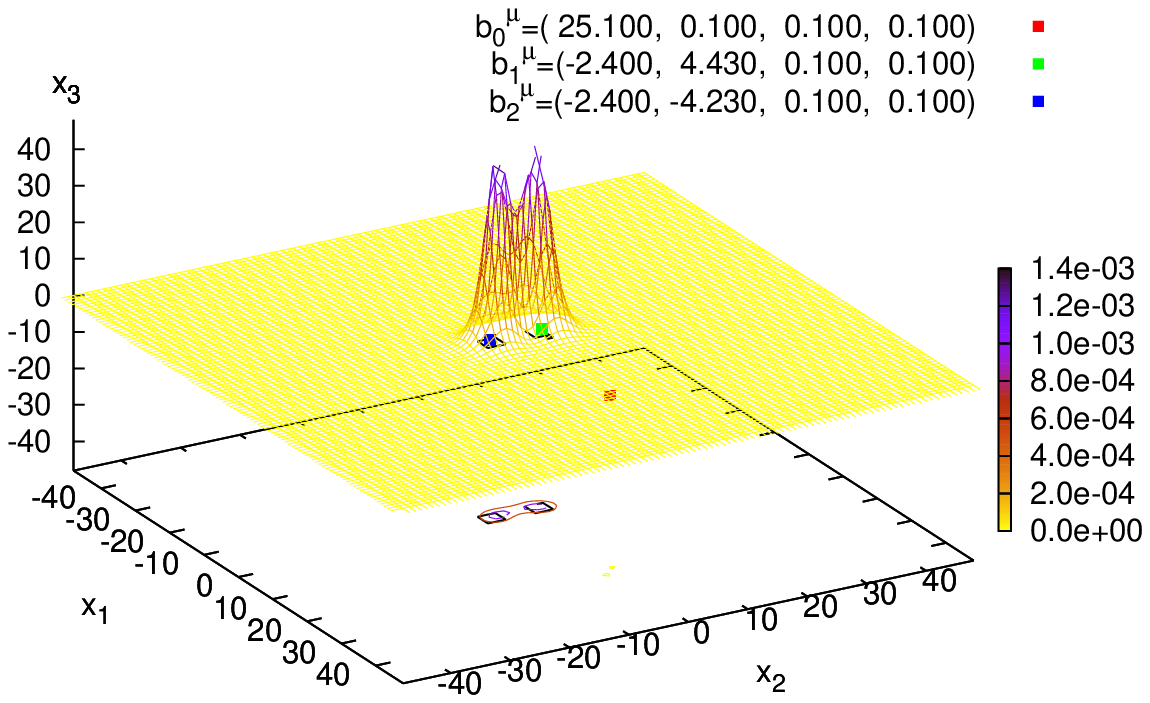}}%
  \put(1400,2250){(e)}
  \put(3500,2250){\includegraphics[trim=0 0 0 0, width=80mm]%
                 {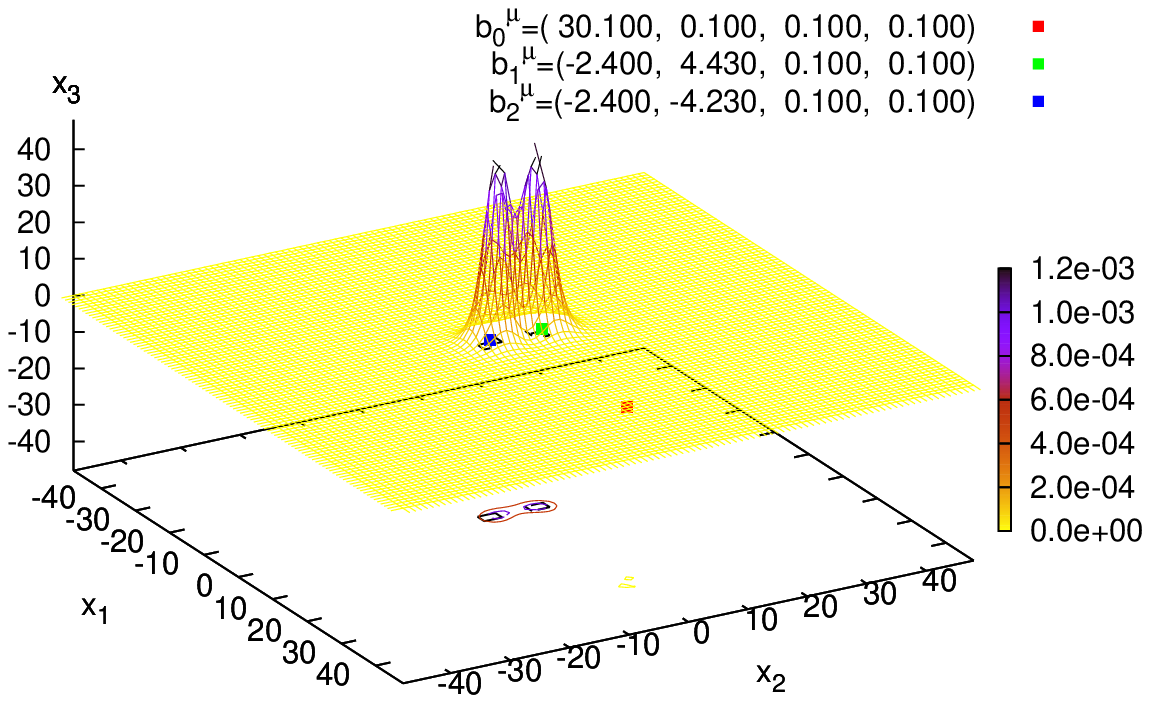}}%
  \put(4900,2250){(f)}
  \put(0,0){\includegraphics[trim=0 0 0 0, width=80mm]%
           {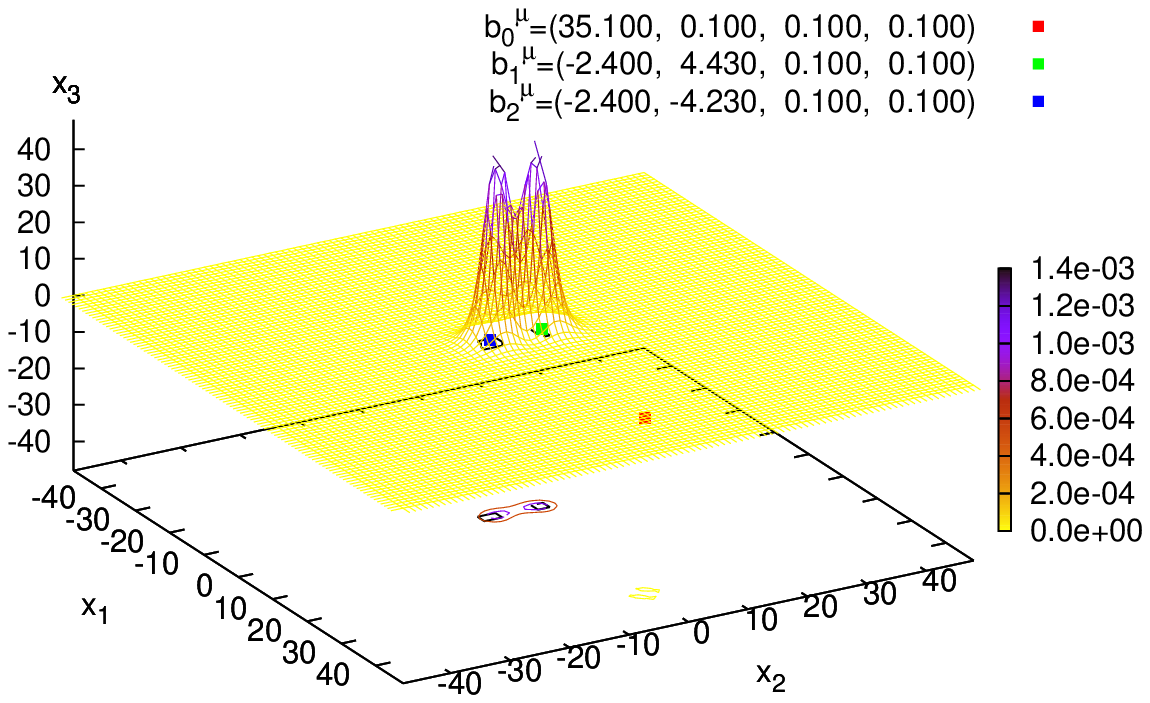}}%
  \put(1400,0){(g)}
  \put(3500,0){\includegraphics[trim=0 0 0 0, width=80mm]%
              {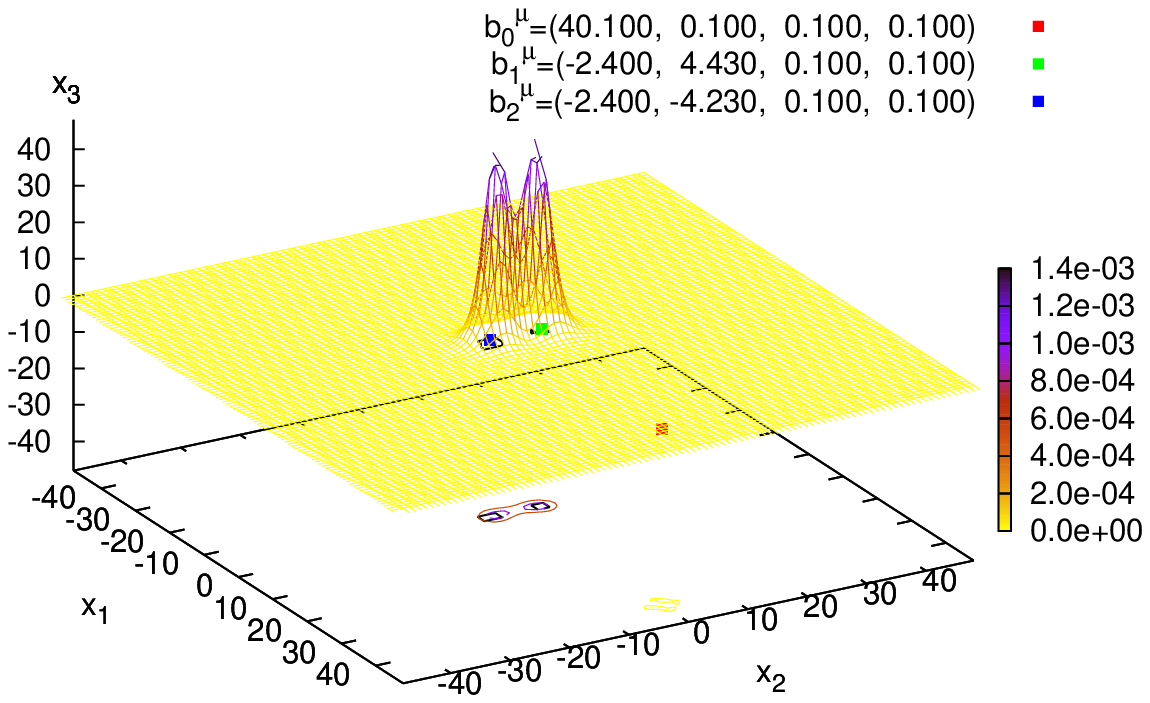}}%
  \put(4900,0){(h)}
 \end{picture}
 \caption{
 The magnetic monopole current $k_{x,\mu}$ generated from the JNR two-instanton
in taking the 't Hooft two-instanton limit.
The parameter  (a) $R=0$, (b) $R=5$, $\dots$, (h)$R=35$ correspond to 
(a) $\rho_0=|b_0|=5$, (b) $10$, $\dots$, (h) $40$.
The grid shows an instanton charge density $D_x$ on $x_1$-$x_2$ ($x_3=x_4=0$) plane.
The black line on the base shows the magnetic monopole loop projected on the $x_1$-$x_2$ plane and the arrow indicates the direction of the monopole current, while colored lines on the base show the contour plot for the equi-$D_x$ lines.
Figures are drawn in units of $a$.
}
 \label{fig:JNRtotHooft_loop}
\end{figure*}

\begin{figure*}[htbp]
 \begin{center}
  \includegraphics[trim=0 130 0 30, width=123mm]{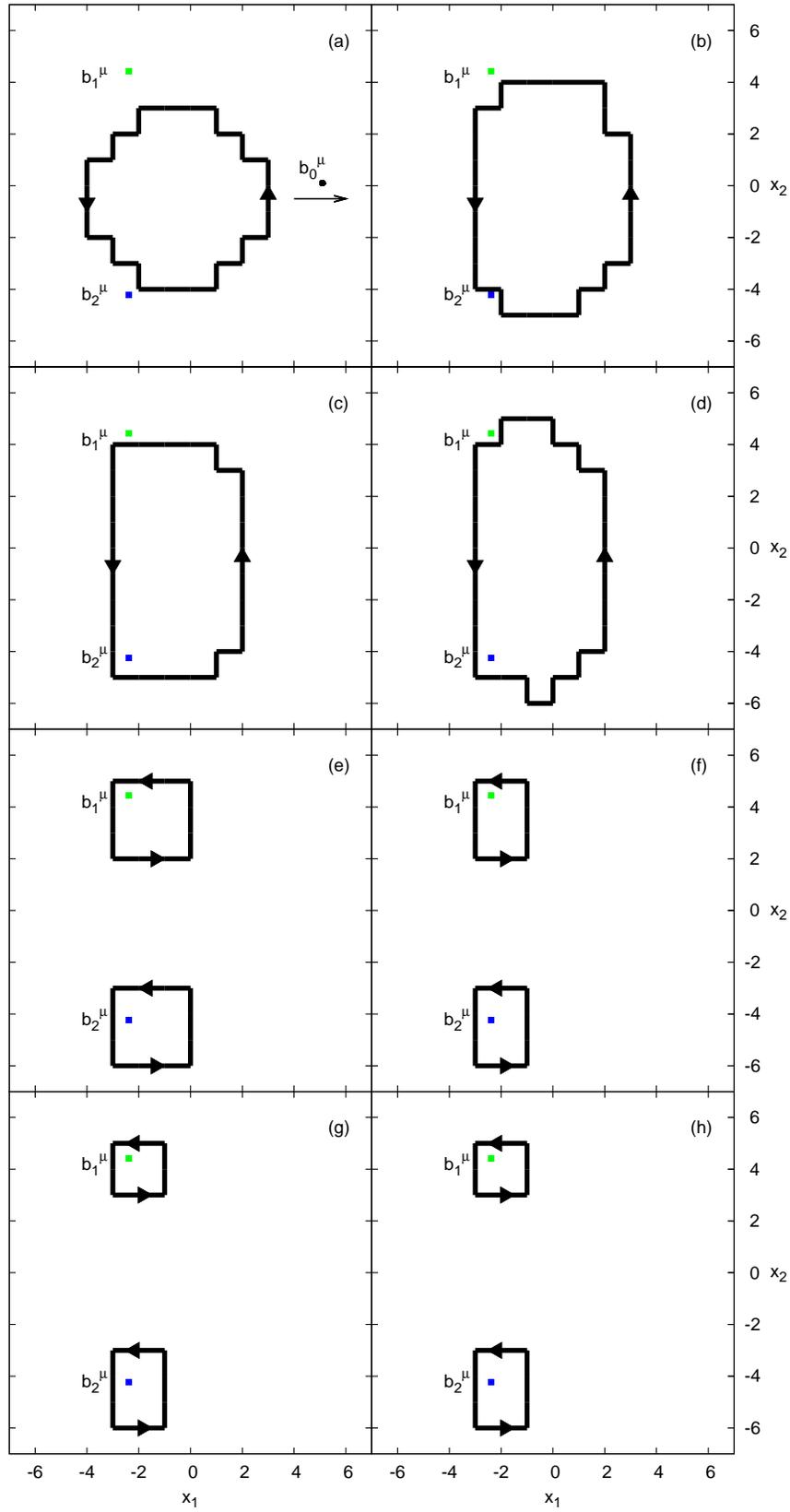}
 \end{center}
 \caption{Deformation of the magnetic monopole loop viewed from the positive side of $x_3$.
The parameter  (a) $R=0$, (b) $R=5$, $\dots$, (h)$R=35$ correspond to 
(a) $\rho_0=|b_0|=5$, (b) $10$, $\dots$, (h) $40$.
As the pole position $b_0^\mu$ shifts to the positive direction of $x_1$ with fixed $b_1^\mu$,
$b_2^\mu$, the magnetic monopole loop shrinks to fixed poles.
The arrow on a loop shows the direction of the monopole current flowing on the loop. 
}
 \label{fig:deformation_loop}
\end{figure*}

\section{Deformation of a magnetic monopole loop generated from the JNR two-instanton}

To investigate how a magnetic monopole loop generated from the JNR
two-instanton behaves in the course of taking the 't Hooft two-instanton limit, 
we set three pole positions and three size parameters to
\begin{align}
 \rho_0=5a+R&,\quad \rho_1=\rho_2 = 5a ,
\\
 b_0=(b_0^1,b_0^2,b_0^3,b_0^4)=& (5a+R,0,0,0)+\Delta ,
\\
 b_1=(b_1^1,b_1^2,b_1^3,b_1^4)
 =& \left(-\frac{5}{2}a,\frac{5\sqrt{3}}{2}a,0,0\right)+\Delta ,
\\
 b_2=(b_2^1,b_2^2,b_2^3,b_2^4)
 =& \left(-\frac{5}{2}a,-\frac{5\sqrt{3}}{2}a,0,0\right)+\Delta,\\
 \Delta=(0.1a,&0.1a,0.1a,0.1a)
\end{align}
and calculate  magnetic monopole current
for various values of $R$: $R=0,5,10,15,20,25,30,35$.
Here we have used the same notations as those used in the previous paper \cite{FKSS10}, in which the details on the lattice, instanton discretization and monopole detection are given.

The results are summarized in FIG.~\ref{fig:JNRtotHooft_loop} and
FIG.~\ref{fig:deformation_loop}.
These figures show that a circular monopole loop generated from the JNR two-instanton splits into two smaller loops as $R$ increases.
Eventually, two smaller loops shrink to two fixed poles in $\rho_0=|b_0|\rightarrow\infty$ limit.
Such behavior of the magnetic monopole loop is consistent with the result  for the 't Hooft two-instanton obtained in \cite{BOT97}.
Therefore, our result on the JNR two-instanton \cite{FKSS10} does not contradict with the preceding result on the 't Hooft two-instanton \cite{BOT97}.

In FIG.~\ref{fig:deformation_loop}, the separation of a loop into two smaller loops occurs between (d) and (e).
The evolution from (d) to (e) seems to be smooth.  
In order to consider this issue, the direction of magnetic current is indicated by the arrow on the loop in Fig.2. 
The directions of magnetic currents of two smaller loops is the same as that of the original loop, which is consistent with the   smoothness of the evolution. 
However, we cannot verify the smoothness by giving the value of the functional to be minimized, since it is difficult to obtain the data at the instance that a loop was just separated into two smaller loops.

The deformation of the magnetic monopole loop seen in FIG.~\ref{fig:deformation_loop}
agrees very well with the one shown by S. Kim and K. Lee analytically
in $(4+1)$ dimensional supersymmetric Yang-Mills theory \cite{KimLee}.

\section{conclusion}

We have investigated numerically how a magnetic monopole loop 
generated from the JNR two-instanton deforms in the course of taking the 't Hooft instanton limit.
In this limit, a circular loop splits into two smaller loops and each loop shrinks to two poles of the 't Hooft instanton. 
This corresponds to the magnetic monopoles demonstrated by  Brower, Orginos and Tan  \cite{BOT97} for the 't Hooft instanton. 
As a result, our previous result \cite{FKSS10} that the JNR two-instanton generates a magnetic monopole loop is compatible with the result \cite{BOT97} that the 't Hooft two-instanton does not generate such a loop of magnetic monopole. 
Thus, there is no contradiction between two results.

 A  natural explanation for the difference between the two instantons' monopoles
  will be possible from the fact that in the 't Hooft ansatz the constituents are of same color
  orientation, whereas in the JNR ansatz they are not \cite{JNR77}, and apparently the latter
 is needed to generate a loop of finite size.
 We hope to clarify this issue in future works.

{\it Acknowledgements}\ ---
The authors would like to thank the referee for carefully reading the manuscript
and giving valuable suggestions for revisions.
This work is  supported by Grant-in-Aid for Scientific Research (C) 21540256 from Japan Society for the Promotion of Science (JSPS), and also in part by the JSPS Grant-in-Aid for Scientific Research (S) \#22224003. 
The numerical calculations are supported by the Large Scale Simulation Program No.09-15 (FY2009) and No.10-13 (FY2010) of High Energy Accelerator Research Organization (KEK).


\end{document}